\documentclass[useAMS,usenatbib]{mn2e}
\usepackage{graphicx}

\def\pmb#1{\setbox0=\hbox{$#1$}
  \kern-0.25em\copy0\kern-\wd0
  \kern.05em\copy0\kern-\wd0
  \kern-0.025em\raise.0433em\box0}
\def\bL{\,\pmb{\mit L}}
\def\br{\,\pmb{\mit r}}
\def\bv{\,\pmb{\mit v}}
\def\bx{\,\pmb{\mit x}}
\def\bz{\,\pmb{\mit z}}

\def\binom#1#2{\left(\kern-.5em\begin{array}{cc} #1 \\ 
                     #2 \end{array}\kern-.5em\right)}

\def\etal{et al.}

\title[Self-similar triaxial halos]{Made-To-Measure Models of
  Self-Similar Triaxial Halos with Steep Inner Density Gradients}
\author[J. C. Malvido \&
  J. A. Sellwood]{J. C. Malvido,$^1$\thanks{E-mail:
    jmalvido@verizon.net} J. A. Sellwood$^2$\thanks{E-mail:
    sellwood@rutgers.edu} \\
  $^1$406 North Road, Chester, NJ 07930, US \\
  $^2$Rutgers University, Department of Physics \& Astronomy, 136
  Frelinghuysen Road, Piscataway, NJ 08854-8019, US}

\pagerange{\pageref{firstpage}--\pageref{lastpage}} \pubyear{2014}

\def\LaTeX{L\kern-.36em\raise.3ex\hbox{a}\kern-.15em
    T\kern-.1667em\lower.7ex\hbox{E}\kern-.125emX}

\begin{document}

\label{firstpage}

\maketitle

\begin{abstract}
We use the Made-to-Measure method to construct $N$-body realizations
of self-similar, triaxial ellipsoidal halos having cosmologically
realistic density profiles.  Our implementation parallels previous
work with a few numerical refinements, but we show that orbital
averaging is an intrinsic feature of the force of change equation and
argue that additional averaging or smoothing schemes are redundant.
We present models having the Einasto radial mass profile that range
from prolate to strongly triaxial.  We use a least-squares polynomial
fit to the expansion coefficients to obtain an analytical
representation of the particle density from which we derive density
contours and eccentricity profiles more efficiently than by the usual
particle smoothing techniques.  We show that our $N$-body realizations
both retain their shape in unconstrained evolution and recover it
after large amplitude perturbations.
\end{abstract}


\section{Introduction}
Dark-matter halos that form hierarchically in simulations of
$\Lambda$CDM cosmology have a number of regular properties:  Their
mass density profiles, the distributions of velocity anisotropy, spin
parameters, and substructure are largely independent of the halo's
mass over a wide range of values.  Other properties, such as the
concentration and triaxiality, vary systematically with halo mass
\citep{Ze09, DM11, FW12}.

The landmark work of \citet[][hereafter NFW]{NFW96} reported that
the halo mass density could be fitted with a broken power-law
\begin{equation}
\rho(r) \propto \left( \frac{r}{r_s}\right)^{-\gamma}
\left( 1+\frac{r}{r_s}\right) ^{\gamma-3}
\end{equation}
where the length scale, $r_s$, is the radius around which the
logarithmic slope changes.  The steepness of the inner cusp reflected
in the value of the exponent $\gamma$ has been the subject of much
debate \citep{Mo99, FM01, Po03, Di04, Di05}.  More recent work
\citep{Na04, Na10, Me05, Ga08} finds the Einasto profile (described in
\S4) with its steep but finite central density to be a better fit.

Virialized halos that have undergone multiple mergers acquire a
triaxial shape in general, though with a strong bias towards
prolateness \citep[]{Fr88, CL96, BS05}.  To accommodate triaxiality
when fitting the mass density, a number of authors \citep[][hereafter
  JS02]{Ka91, DC91, JS02} replace the radial coordinate $r$ with the
dimensionless coordinate $\tilde\xi$ defined through
\begin{equation}
\tilde\xi^2 = \frac{x^2}{a^2} + \frac{y^2}{b^2} + \frac{z^2}{c^2}\quad
,\quad a\geq b\geq c
\label{homoeoidal surface}
\end{equation}
where $a$, $b$, and $c$ are the lengths of the major, intermediate,
and minor axes, respectively.  Although axial ratios and alignments
display some radial variation \citep[JS02;][hereafter SFC12]{CL96,
  BS05, SFC12}, we focus here on creating self-similar ellipsoids for
which these values remain in a fixed ratio, with the axes aligned, at
all radii.  This idealized geometry reduces the mass density to a
function of a single variable $\tilde \xi$ (eq.~\ref{homoeoidal
  surface}), with the axis ratios $b/a$ and $c/a$ as constant
parameters.

Self-similarity is an attractive assumption, but it is generally
insufficient to enable a derivation of a distribution function (DF).
The DF can be expressed in terms of the actions \citep{BT08}, but
these are known only in the case of the ``perfect'' ellipsoid for
which the potential is fully separable in ellipsoidal coordinates
\citep{dZLB85, dZ85} and all orbits are regular.  Regular orbits also
exist in more general ellipsoids but irregular orbits, for which the
only isolating integral is the energy, populate some parts of phase
space, and may even dominate \citep{UP88, VM98, Va10}.

Various approaches have been adopted to construct $N$-body models of
triaxial mass distributions with some success.  \citet{Ho01} and
\citet{Wi08} ``squeeze'' a spherical model, \citet{RS06,RS09}
developed an iterative method that corrects the density after each
time step, without revising the velocities, until equlibrium is
attained, while \citet{Mo04} construct them through multiple
collisions of separate systems.  However, neither the squeeze nor
collision methods provide a systematic means for reproducing a
specific geometry, such as self-similarity, while only halos with a
central core, rather than a steep inner density gradient, have been
attempted with the iterative method.

\citet{Sch79, Sch93} built triaxial models from linear combinations of
bound orbits in the desired potential.  In his method, orbits are
drawn from a pre-calculated library that approximately spans the
phase-space of the bound system and contains the time-averaged
contributions to the mass density in a set of spatial cells.  The
assignment of orbital weights is constrained by the targeted mass
density in these cells, but since the assignment is far from unique,
it is customary to seek an optimal solution, such as by maximizing the
entropy \citep{Ri88} to favor a smooth distribution.  The method has
been successfully applied to cuspy, triaxial systems \citep{MF96,
  Me97, Ca07, VZ08, VA12}.  However, creation of $N$-body models
requires the further, non-trivial step of selecting particles from
appropriately chosen points along the orbits selected from the
library.

Here we adopt the ``made to measure'' (hereafter M2M) algorithm
proposed by \citet[][hereafter ST96]{ST96}, which is similar in spirit
to Schwarzschild's but adjusts the mass of each moving particle
dynamically.  The algorithm has been further developed by
\citet[][hereafter DL07]{DL07}, \citet[][hereafter D09]{De09},
\citet{LM10}, \citet{MG12}, and \citet{HK12}, and a useful brief
review is given by \citet{Ge10}.  A further advantage of M2M is that,
on completion, the equilibrium $N$-body system is ready to use with no
additional step.  The range of applications of M2M \citep{DL08, DL09,
  Da11, LM12, Lo12} is a fair measure of its versatility.

One can add observational constraints to both M2M (DL07)
and Schwarzschild's method \citep[see e.g.][and references
  therein]{Ch08} in order to construct models for comparison with
data, which is probably the most widely used application of these
methods.  But our objective in this paper is purely theoretical.

Our aim is to use M2M to construct $N$-body realizations of
self-similar ellipsoidal halos with a mass density and
eccentricity suggested by cosmological simulations.  While an
improvement on spherical halos, our models are still idealizations,
since cosmological halos are generally not self-similar, and we also
neglect rotation \citep[e.g.][]{Bu01}, certain kinematic features
\citep{CL96, Wo05, Ha09} and substructure \citep{Di04, Di05, Sp08}.

Some triaxial models have already been created by this technique, notably
in DL07 and D09, but there are minor differences in our approach and
we present more extensive tests of the stability of the resulting
models.  We will use them in later work to study the influence of halo
triaxiality on the dynamics of embedded disks.

\section{The algorithm}
The M2M method has already been described elsewhere (e.g., ST96, DL07,
D09).  We therefore give only an outline, noting where our treatment
differs from previous work.  Our presentation follows closely that
given by DL07, but we do not require their generalization to include
constraints from velocity data.

\subsection{Preliminaries}
\label{M2M prelim}
We seek to create equilibrium halo models with a target density
distribution $\rho_T(\bx)$ and total mass ${\cal M}_T$ in the
gravitational potential $\Phi_T(\bx)$.  In most applications, the
potential arises from the density $\rho_T$ only, but it is also
possible to consider models with additional mass components, such as a
disk or central mass, that contribute to $\Phi_T$ but not to $\rho_T$.

First we divide the volume of the target model into a set of $K$ cells, each
having a volume ${\cal V}_k$, and define a set of occupancy kernels
$\left\{{\cal K}_{k}(\bx), \;k = 1, \ldots, K\right\}$
\begin{equation}
{\cal K}_k\left( \bx\right) =\cases{1 & $\bx$ is within ${\cal V}_k$ \cr 
0 & otherwise.}
\label{Cell-occupation Kernel}
\end{equation}
The mass in the $k$th cell is therefore given by
\begin{equation}
P_k = \int {\cal K}_k(\bx) \rho_T(\bx)\, d^3\bx.
\label{Target moments}
\end{equation}

Suppose that we were to construct a trial set of particles (e.g., as
described in \S\ref{Initialization})
\begin{equation}
{\cal F}_0(\bz) = m_p\sum_{i=1}^{N}w_{0i}\delta(\bz-\bz_i(t=0)),
\label{Trial distribution}
\end{equation}
where $\bz = (\bx,\bv)$ is a 6D phase space coordinate.  The initial
density is determined by the adopted set of $N$ particle position
coordinates $\{\bx_i(0)\}$ with their initial fractional weights
$\{w_{0i}\}$ normalized by the fiducial particle mass $m_p$ (see
\S\ref{Particle Resolution}) such that $m_p\sum_{i=1}^{N}w_{0i} ={\cal
  M}_T$.  At $t=0$, this initial guess therefore has the mass in each
of $K$ cells
\begin{equation}
p_{0k}= m_p\sum_{i=1}^N w_{0i} {\cal K}_k(\bx_i(0)).
\end{equation}
The density at later times, $\rho(\bx,t)$, is determined by
integrating all particle trajectories to time $t$ in the frozen
potential $\Phi_T$, when the mass in each cell becomes
\begin{equation}
p_k(t) = m_p\sum_{i=1}^Nw_i(t){\cal K}_k(\bx_i(t)).
\label{Sampled moments}
\end{equation}

Schwarzschild's approach is first to compute the time average
occupancy of each particle in each cell
\begin{equation}
{\cal K}_{ik}\equiv \frac{1}{\tau}\int_0^\tau {\cal K}_k(\bx_i(t))\, dt,
\label{Orbital average}
\end{equation}
for some large but finite time period $\tau$, and then to require that 
\begin{equation}
\langle p_k\rangle = m_p\sum_{i=1}^N w_i {\cal K}_{ik} = P_k,
\label{Schwarzschild condition}
\end{equation}
for a set of time-independent and non-negative $\{w_i\}$.  Since the
space of possible solutions to eq.~(\ref{Schwarzschild condition}) is
generally large, it is desirable to optimize for smoothess; a popular
choice is to maximize the ``entropy'', which is defined as
\begin{equation}
{\cal S} = -\frac{1}{N}\sum_{i=1}^Nw_i\ln \left[\frac{w_i}{\tilde w_i}\right].
\label{Entropy}
\end{equation}
We use our initially guessed weights for the priors, i.e.\ $\tilde w_i
= w_{0i}$, but other choices are possible.  An entropy decrease,
$\Delta {\cal S} <0$, may be interpreted as an increase of ``order''
in the physical system; for example, a sphere is less ordered than an
ellipsoid.

\subsection{Made to measure}
\label{M2M averages}
In the M2M method, the particle weights $w_i(t)$ are adjusted, while
maintaining the condition $\rho({\bx},t) \approx \rho_T({\bx})$, as
the particle trajectories are integrated.  It is usually cast as an
optimization problem in which the ``likelihood function''
\begin{equation}
{\cal L}(t) =\mu{\cal S}(t) - {\cal C}(t) , \quad\hbox{with}\quad \mu >0
\label{likelihood function}
\end{equation}
is to be maximized.  Both parts of this objective are time-dependent,
since the weights are adjusted.  The ``cost function'' is
\begin{equation}
{\cal C}(t) = \frac{1}{2}\sum_{k=1}^K
\left[\frac{p_k(t)-P_k}{\hat\sigma_k}\right]^2,
\label{chi-squared statistic}
\end{equation}
where $\hat\sigma_k$ is the dispersion of $p_{0k}$, i.e., the sampling
uncertainty implicit in generating ${\cal F}_0(\bz)$ (DL07, D09).

The condition for the optimal solution for $\{w_i^\ast\}$ when
$w^\ast_i \geq 0$ is \citep{Av76}
\begin{equation}
w^\ast_i \frac{\partial{\cal L}}{\partial w_i} = 0. \quad i=1,\dots,N
\label{Kuhn-Tucker}
\end{equation}
Therefore, either $w_i^\ast = 0$, or
\begin{equation}
\frac{\partial {\cal L}}{\partial w_i} = \mu \frac{\partial S}{\partial w_i}
 - \frac{1}{N} \sum_{k=1}^K\frac{{\cal K}_k(\bx_i)}{\hat\sigma_k}\Delta_k = 0,
%\quad i=1,\ldots ,N
\label{Extremum condition}
\end{equation}
where $\Delta_k\equiv (p_k-P_k) /\hat\sigma_k$.

\subsection{The FOCE}
\label{sec:FOCE}
Originally proposed by ST96, the force-of-change-equation (FOCE)
\begin{equation}
\dot w_i = \epsilon w_i \frac{\partial {\cal L}}{\partial w_i},
\label{FOCE}
\end{equation}
with $\epsilon$ being a small parameter, provides a rule for adjusting
$w_i$ in order to drive $\partial {\cal L}/\partial w_i$ towards zero,
a procedure in a simulation that is readily combined with the particle
motion.

The $w_i(t)$ and $\Delta_k(t)$ functions have noisy paths owing to
$N$-body sampling noise and the movement of particles across cells,
which causes abrupt changes to the values of ${\cal K}_k(\bx_i)$.
ST96 suggested replacing the $\Delta_k$ in eq.~(\ref{FOCE}) with its
{\it moving average}, $\tilde\Delta_k$, such that $\tilde\Delta_k(0)
= \Delta_k(0)$ and
\begin{equation}
\frac{d\tilde\Delta_k}{dt} = \eta \left( \Delta_k - \tilde\Delta_k \right)
\label{ODE for averaged deltas}
\end{equation}
where $\eta \geq 2\epsilon $, as a means of suppressing large
fluctuations in the FOCE from these sources of noise and likened its
effect to that of smearing a set of virtual particles over a given
orbit.  This adjustment has been integral to implementations of M2M
reported in the literature.  D09 points out, however, that this
procedure is inconsistent with the extremum condition
(\ref{Kuhn-Tucker}) and compromises the meaning of $\cal C$ in
eq.~(\ref{chi-squared statistic}) since the $\hat\sigma_k$ reflect the
original $N$-body shot noise, not the reduced noise.  Instead, D09
applies a moving average to $\partial {\cal L}/\partial w_i$, and, in
combination with eq. (\ref {FOCE}), obtains a second order ODE for
$\ln w_i$.

However, the following argument suggests that such attempts at
explicit averaging are redundant.  Formally, the FOCE may be
integrated to give
\begin{equation}
w_i(t) = e^{\epsilon D_i(t)}w_{i0},
\label{Integral equation}
\end{equation}
where 
\begin{equation}
D_i(t) \equiv \int_0^t {\partial{\cal L} \over \partial w_i}\, dt^\prime.
\label{Evolution operator}
\end{equation}
Thus we see that $\epsilon D_i(t)$ is proportional to the time average
$\langle \partial {\cal L}/\partial w_i\rangle_t$.  Substitution of
eq.~(\ref{Extremum condition}) for the integrand in (\ref{Evolution
  operator}) shows that the behavior of $\epsilon D_i(t)$ is dominated
by orbital averages of the form ${\left\langle {\cal
    K}_k(\bx_i)\Delta_k\right\rangle}_t$.  But ${\cal K}_k(\bx_i)$ is
influenced by the trajectory of the $i$th particle only, while
$\Delta_k$ is determined by all particles in the $k$th cell.  If the
number of such particles is sufficiently large, it is reasonable to
expect that ${\left\langle {\cal K}_k(\bx_i) \Delta_k\right\rangle}_t$
decouples at large times into the product of two independent averages
${\cal K}_{ik} \times {\left\langle \Delta_k\right\rangle}_t$, where
${\cal K}_{ik}$ is defined in eq.~(\ref{Orbital average}).
Furthermore, ST96 and DL07 prove that ${\left\langle
  \Delta_k\right\rangle}_t \rightarrow 0$ at large times which, from
the definition of $\Delta_k$, implies $\langle p_k\rangle_t
\rightarrow P_k$.  Thus we effectively recover eq.~(\ref{Schwarzschild
  condition}) and thereby formally establish the asymptotic
equivalence of the M2M and Schwarzschild's methods.

Therefore, not only is explicit averaging redundant but the
application of eq.~(\ref {ODE for averaged deltas}) is
counterproductive if $N$ is large enough.  We have not explored how
large $N$ needs to be for this statement to be true, but when we
employ $N = 1.2 \times 10^6$ particles, we find that inclusion of the
averaging procedure (\ref{ODE for averaged deltas}) degrades the
actual value of $\cal C$ by as much as an order of magnitude.  We
therefore refrain from including any additional averaging.

Thus we find a deeper resemblance between the M2M approach and
Schwarzschild's method.  However, there are two reasons for preferring
M2M: not only does eq.~(\ref{Extremum condition}) combine averaging
and solving for $\left\{w_i^\ast\right\}$ into a single algorithm
instead of two distinct steps but, upon convergence, it also delivers
a representative set of particle coordinates that are in equilibrium,
in readiness for an $N$-body simulation.

\subsection{Convergence}
Fluctuations in the values of the $w_i(t)$ as the system evolves
jointly with the FOCE imply that satisfying condition
(\ref{Kuhn-Tucker}) for all $N$ particles simultaneously at any given
instance is implausible.  Instead, an equally effective, yet more
realistic, goal of the M2M method is to ensure that $\max\Vert \langle
\Delta_k \rangle \Vert < \delta$, for some prescribed threshold
$\delta$.

By linearizing the FOCE (exclusive of the entropy term), ST96 (see
also DL07) obtained a set of equations for the modes of the $\langle
\Delta_k\rangle$ and demonstrated that these modes decay exponentially
-- a result that essentially formalizes the intuition behind the
intent of the FOCE.  Though it does not provide practical guidelines
for assessing convergence, it does, however, render the $\langle
\Delta_k\rangle$, rather than the $w_i(t)$, as the focus of the
convergence analysis.

The threshold, $\delta$, may be made as small as desired by reducing
$\mu$, with the disadvantage that an increasing number of particle
weights are driven toward zero.  Reaching a desired threshold $\delta$
is not necessarily indicative of convergence; only when changes in the
entropy become small relative to its level can the algorithm be said
to have reached convergent conditions.

Ultimately, the true test of the algorithm's convergence is whether
the resulting system maintains the desired shape and mass profile as
it evolves freely and \textit{self-consistently} without intervention
from the FOCE (see \S\ref{sec.stability}).

\subsection{Summary of the Procedure}
Our implementation of the algorithm, which will be discussed in
greater detail later, may be conveniently summarized as follows (see
also D09):

\begin{enumerate}
\item \textit{Initialization}: Construct an equibrium, spherically
  symmetric model with the desired mass density profile consisting of
  $N_M$ particles.  Transform the particle coordinates by rescaling
  (see eq.~\ref{Scale transformation} below) to achieve the targeted
  geometry and adjust the velocities so that the tensor virial theorem
  is satisfied.

\item \textit{Relaxation}: Allow the system to settle (without the
  FOCE) for 20 -- 25 dynamical times.

\item \textit{M2M step}: Evolve the system while adjusting the weights
  in accordance to the FOCE.

\item \textit{Convergence check}: Once the entropy level is nearly constant,
switch off the FOCE and allow the system to evolve
self-consistently to test whether it is in equilibrium.

\item \textit{Stability Check}: Apply an adiabatic, aspherical
  perturbation to seed any possible secular instabilities and
  follow the subsequent self-consistent evolution.
\end{enumerate}

\noindent Note that we obtain a smooth target potential $\Phi_T$ and
the values of $\hat\sigma_k$ needed for steps (ii) and (iii) by
creating a model having many more particles than is practical to
evolve, i.e., $N_T \gg N_M$.  Their positions, with velocities
ignored, give a smoother representation of $\rho_T$, and we can solve
for their gravitational field using our adopted $N$-body method.

It is possible to run M2M using a self-consistent, rather than a
rigid, potential.  This is less satisfactory, however, because M2M
continuously tries to adjust weights to compensate for (mild) particle
scattering.  We therefore prefer to use a smooth and frozen target
potential.

\section{Computational Methods}
\label{Computational Methods}
\subsection{Self-Similar Ellipsoid}
\label{Homoeoid}
The target triaxial halos in this work consist of a series of concentric and
co-eccentric, nested ellipsoidal shells - a simplifying idealization.
Surfaces of constant density in a self-similar ellipsiod have constant
$\xi$, where
\begin{equation}
\xi^2 = x^2 + y^2\frac{a^2}{b^2} + z^2\frac{a^2}{c^2}.
\label{Self-similarity condition}
\end{equation}
Here $a$, $b$, and $c$ are the semi-axes of the outer surface of the
finite ellipsoid.  We also define the eccentricities, $\varepsilon_y =
(1 - b^2/a^2)^{1/2}$ and $\varepsilon_z = (1 - c^2/a^2)^{1/2}$ for
later use.  Strongly triaxial systems have values of the triaxiality
parameter, ${\bf T \equiv (a^2-b^2)/(a^2-c^2)} \sim 0.5$, while $T=0$
for an oblate, and $T=1$ for a prolate, spheroid.

We wish to construct self-similar ellipsiods for which the density
profile $\rho_e(\xi)$ has the same functional form (to within a
normalization factor to preserve the same total mass) as some
spherical model $\rho_s(r)$.  We relate a point $\br = (x,y,z)$ on a
sphere to $\br^\prime = (x^\prime,y^\prime,z^\prime)$ on the
ellipsoidal surface through
\begin{equation}
r^2 = x^2 + y^2 + z^2 = x^{\prime 2} + y^{\prime 2} a^2 / b^2 +
z^{\prime 2} a^2/ c^2 = \xi^2,
\label{Shell mapping}
\end{equation}
which is equivalent to the scale transformation 
\begin{equation}
x^\prime = x \quad ;\quad y^\prime = y a/b \quad ;\quad z^\prime = z a/c.
\label{Scale transformation}
\end{equation}
Two ellipsoidal surfaces at $\xi$ and $\xi + \delta\xi$ bound a volume
known as a {\it thin homoeoid} (BT08, \S2.5).  The volume of a thin
homoeoid is $4\pi \theta_{yz}\xi^2d\xi$, with $\theta_{yz} \equiv
bc/a^2$, while that of a spherical shell is $4\pi r^2dr$.  Therefore,
if a sphere of radius $a$ and density $\rho_s(r)$ is compressed along
the intermediate and minor axes by factors $b/a$ and $c/a$
respectively in such a way that equidensity shells do not cross, it
yields a self-similar ellipsoid of equal mass with density
$\rho_e(\bx) = \theta_{yz}^{-1}\rho_s(\xi)$.

\subsection{Force computation}
Particle-mesh, or grid, methods remain the most efficient for $N$-body
simulations of an isolated gravitating system \citep{Se14}.  Here we
use a radial grid that computes forces from an expansion in surface
harmonics on a set of spherical shells that are more closely spaced
near the center.  An arbitrary field function $\Psi(\bx)$ (e.g., the
gravitational potential, force components, mass density, etc.) can be
written in real form as
\begin{eqnarray}
\Psi(\bx) & = & \sum_{l=0}^\infty\sum_{m=0}^l \gamma_{lm}
\Pi_l^m(\theta) \nonumber \\ && \qquad \qquad \left[ \Psi_{lm}^c(r)\cos
  m\phi + \Psi_{lm}^s(r)\sin m\phi \right],
\label{Harmonic expansion}
\end{eqnarray}
where $\bx = (r,\theta,\phi)$ is the field point, 
\begin{equation}
\Pi_l^m(\theta) \equiv \sqrt{\frac{(l-m)!}{(l+m)!}}P_l^m(\cos\theta),
\label{Normalized Legendre Polynomials}
\end{equation}
$P_l^m(\cos\theta)$ is the associated Legendre function, and the
normalization factor $\gamma_{lm} = (2-\delta_{m0}) (2l+1) /4\pi$.
The coefficients in eq.~(\ref{Harmonic expansion}) can be computed from
\begin{equation}
\binom{\Psi_{lm}^c(r)}{\Psi_{lm}^s(r)} = \int \Pi_l^m(\theta^\prime)
\binom{\cos m\phi^\prime}{\sin m\phi^\prime}\Psi(r,\theta^\prime,\phi^\prime) 
\sin\theta^\prime \; d\theta^\prime d\phi^\prime.
\label{Harmonic components}
\end{equation}
Truncating the expansion (\ref{Harmonic expansion}) at some $l =
l_{\rm max}$ smooths small-scale angular fluctuations.

We employ the $N$-body scheme described by \citet{Se03} that uses a
1D grid for the radial coordinate while retaining the exact angular
expansion for each particle.  The radial nodes are spaced
logarithmically, and subdivide a spherical volume into shells of
gradually increasing thickness.  The functions $\Psi^\alpha_{lm}$ are
tabulated at shell boundaries and evaluated at other radii by linear
interpolation.  To ensure radial continuity, individual particle
masses are divided between neighboring grid points according to the
cloud-in-cell procedure, which also distinguishes between mass
interior and exterior to a field point.  The particles are smeared in
angle because the expansion (\ref{Harmonic expansion}) is limited by
the adopted $l_{\rm max}$.  The end result may be pictured as each
point particle being replaced by a smeared density distribution wedged
between neighboring spherical radial shells.  See \citet{Se03} or {\tt
  http://www.physics.rutgers.edu/$\sim$sellwood/manual.pdf} for a full
description.

\subsection{M2M Binning}
A particle within the $k$th
radial shell, which spans the range $(r_{k-1},r_k)$, contributes to
$(l_{\rm max}+2)(l_{\rm max}+1)$ real terms $\Psi_{lm}^\alpha(r_k)$.
If we were to use the gravity grid shells to define the cells in
eq.~(\ref{Cell-occupation Kernel}) as in DL07, then 
$n_r+1$ radial nodes would yield a total of $n_r
(l_{\rm max}+2) (l_{\rm max}+1)$ separate kernel moments ${\cal
  K}_{k;lm}^\alpha$ (eq.~\ref{harmonic kernel}); note that each
particle contributes to a single radial shell $k$ and to multiple
$(l,m,\alpha)$ components.

Since $n_r$ typically ranges in the hundreds, we superimpose a
coarser grid -- the M2M grid -- consisting of $k_r+1 \ll n_r$ nodes,
where all nodes in the M2M grid coincide with nodes of the finer
gravity grid.  Coarser radial spacing not only reduces the particle
shot noise in each bin but also improves the efficiency in solving the
FOCE by reducing the number of kernel moments to be computed.

It may seem that one of these grids is superfluous, for if the
resolution of the M2M grid is adequate to resolve the mass density,
then it should be adquate to resolve the gravitational forces.
However, as argued above, the FOCE averages over a large number of
particle trajectories making coarser M2M binning acceptable, while the
finer gravity-grid is needed to resolve forces and determine accurate
particle trajectories.

\subsection{Unequal particle masses}
\label{Particle Resolution}
Even though they are less closely spaced than the gravity nodes, the
M2M grid spacings should still be small, especially in the inner half
of the halo, in order to represent the targeted shape.  But with equal
mass particles, fine binning causes larger statistical fluctuations in
the cell occupancy.  We therefore employ particles
that are less massive near the center and have gradually increasing
masses towards the outer edge.  To achieve this, we use the
dimensionless function $W(L) = L_0 + L$ to weight the {\it initial}
particle masses.  Here, $L\;(\equiv|\bL|)$ is the specific angular
momentum (in our dimensionless units, see \S\ref{sec.units}) and $L_0$
a convenient constant.  To maintain the desired initial density, we
select particles from the DF weighted by $W^{-1}(L)$ to ensure that
lighter particles are proportionately more numerous \citep{Se08}.

We store the initial value $w_{i0} = W(L_i)$ for each particle, which
we subsequently adjust during M2M as required by the FOCE
(eq.~\ref{FOCE}).  Furthermore, we adopt the initial weights $w_{i0}$
for the priors $\tilde w_i$ in the entropy term (\ref{Entropy}).

Since ${\cal M}_T = \int\int f \,d^3\bv d^3\bx$, the fiducial particle
mass
\begin{equation}
m_p = {1 \over N}\int\int {f \over W} \,d^3\bv d^3\bx,
\label{mean mass}
\end{equation}
so that the actual mass of each particle is $m_p$ times its current
weight $w_i(t)$, and $m_p\sum_i w_{i0} \equiv {\cal M}_T$.

\subsection{Target Moments and Kernel Computation}
\label{Density Fits}
Estimates of the local density from a system of $N$ particles are
notoriously noisy, so we work with integrated masses.
As in DL07, we define the $(l,m,\alpha)$ \textit{mass harmonic}
component in the $k$th bin of the M2M grid as
\begin{equation}
M_{k;lm}^\alpha = \int_{r_{k-1}}^{r_k}r^2\rho_{lm}^\alpha(r)\,dr,
\label{Theoretical binned mass}
\end{equation}
where values of $\rho_{lm}^\alpha(r)$ are derived from
eq.~(\ref{Harmonic components}) with $\Psi(\bx)$ replaced by
$\rho_T(\bx)$.

Applying eq.~(\ref{Theoretical binned mass}) to a discrete particle
density yields the corresponding harmonic mass components $m_{k;lm}^\alpha$
due to particles
\begin{equation}
\binom{m_{k;lm}^{c}}{m_{k;lm}^{s}} = m_p\sum_{i, k}^Nw_i\Pi_l^m(\theta_i)
 \binom{\cos m\phi_i}{\sin m\phi_i},
\label{Particle binned mass}
\end{equation}
where $m_p$ is given by eq.~(\ref{mean mass}).  Similarly, the
kernel ${\cal K}_{k}(\bx_i)$ in eq.~(\ref{Cell-occupation Kernel}) is given by
\begin{equation}
\binom{{\cal K}_{k;lm}^c(\bx_i)}{{\cal K}_{k;lm}^s(\bx_i)} = 
\cases{m_p\Pi_l^m(\theta_i) \binom{\cos m\phi_i}{\sin m\phi_i} &
$\bx_i$ in bin $k$ \cr
0 & otherwise.}
\label{harmonic kernel}
\end{equation}
For the FOCE, the deviation becomes
\begin{equation}
\Delta_{k;lm}^\alpha = \frac{m_{k;lm}^\alpha - M_{k;lm}^\alpha}{\sigma_{k;lm}^\alpha}.
\label{harmonic delta}
\end{equation}
\S3.9 describes how we estimate the denominator $\sigma_{k;lm}^\alpha$.

\subsection{The FOCE}
We use a forward difference to integrate the FOCE, a first order ODE.
The parameter $\epsilon$ determines the rate of change of $\dot{w}_i$,
which is also driven by contributions from all other particles in the
same radial bin.  By iterating multiple $n_F$ times at fixed $t$ with
the adjusted constant $\epsilon/n_F$ until the solution of the FOCE is
well converged, we improve the overall convergence properties of the
algorithm and diminish the fluctuations in ${\Delta}_k$.  After
each intermediate iteration, we update the right side of the FOCE and
renormalize to preserve the total mass $m_p\sum_iw_i(t)$.

DL07 suggest the value of $\epsilon$ should be a small parameter,
$\epsilon_0$, times $[\max_{i,k}\vert {\cal K}_k(i) \Delta_k /
  \sigma_{k} \vert]^{-1}$.  We adopt their suggestion, but since the
maximum value is subject to large swings, especially at the outset, we
limit its time variability by employing a moving average.

\subsection{Halo Initialization}
\label{Initialization}
Our method to prepare the model for M2M integration is similar to that
described by D09, whereas DL07 employ the squeeze method to obtain the
initial equilibrium, triaxial figure.  We proceed as follows:
\begin{enumerate}
\item Determine the DF for a spherically symmetric, isotropic halo
  of density $\rho_s(r)$ by Eddington inversion, and then redefine the
  DF to be zero for $E > E_{\rm max}$, where $E_{\rm max} =
  \Phi(r_{\rm max})$, with $r_{\rm max} = a$, the semi-major axis of
  the target ellipsoid.  The density tapers smoothly to zero at
  $r_{\rm max}$, even though the truncation in $E$ is sharp.  We apply
  the weighting scheme described in \S\ref{Particle Resolution} and
  select $N_M$ particles in the usual way \citep[][\S2.3]{SD09}.  This
  procedure yields a self-consistent spherical model that is close to
  equilibrium; the change in central attraction near the outer edge
  caused by the tapered density creates only a very mild imbalance.

\item Apply the scaling transformation, eq.~(\ref{Scale
  transformation}), to the spatial coordinates of the particle
  phase-space generated in step (i) in order to deform the sphere onto
  a self-similar ellipsoid, without changing the particle velocities.

\item Evaluate the potential energy $W_{ij}$ and kinetic energy
  $K_{ij}$ tensor components using the {\it frozen} gravitational
  field of the target ellipsoid (determined as described below) and
  velocities from the spherical equilibrium model, respectively, and
  rotate the phase-space coordinates slightly so that the kinetic and
  potential energy tensors become rigorously diagonal.  We also adjust
  the velocity components of each particle so that on average
  $W_{ii}/2K_{ii} =  1$ for all three Cartesian components.

\item Evolve the system for 25 dynamical times, defined in
  \S\ref{sec.units} below, in the fixed gravitational field of the
  target ellipsoid to allow for relaxation during which the model
  becomes slightly less flattened.
\end{enumerate}

\subsection{M2M grid}
In determining the nodes of the M2M grid, we use the spherical radii
of the $N_M$ particles generated in step (i) above to ensure that each
radial bin initially contains at least a preset number of particles,
or extends over a maximum prescribed number of gravity grid nodes,
whichever criterion yields the fewer nodes.  Thus, bins near the
center where the density is high have the radial extent of only a few
gravity grid nodes, while bins are as large as we allow in lower
density regions.

\subsection{Target Ellipsoid}
To construct a {\it target potential} that is as smooth as
possible, we duplicate steps (i) and (ii) of the above halo
initialization procedure for a system of $N_T$ particles, with $N_T
\gg N_M$.  The target potential is that of the larger number of
particles, which suffers from milder shot noise.

Moreover, we draw multiple, independent samples of $N_M$ particles
each from the $N_T$ target population in order to estimate the mean
target moments, $\bar M_{k,lm}^\alpha$, and associated dispersions,
$\sigma_{k;lm}^\alpha$ (eq.~\ref{harmonic delta}), based on the
statistical spreads of the samples.

\section{An Example: The Einasto Halo}
\subsection{Mass Density}
\citet{Na04, Na10}, \citet{Me05, Me06}, and \citet{Ga08} show that the
logarithimc slope of the spherically-averaged mass density in
well-resolved cosmlogical halos can be fitted as a power law,
$d\ln\rho(r)/d\ln r = -2(r/r_s)^\kappa$, with the value of $\kappa$
dependent on the mass of the halo.  The scale $r_s$ is the radius at
which the logarithmic slope $= -2$.  The density that has this
property is the Einasto profile \citep{EH89}
\begin{equation}
\rho_{\rm E}(r) =\frac{M_0}{16\pi r_s^3}\exp \left\{ -\frac{2}{\kappa
}\left[ \left( \frac{r}{r_s}\right)^\kappa-1\right] \right\},
\label{Einasto formula}
\end{equation}
which rises steeply towards the center, but remains finite.
\citet{Ca05} give formulas for the mass, potential, and force in terms
of the incomplete gamma function $\gamma(p,x)$ and its complement
$\Gamma(p,x)$, which is distinguished from the usual Gamma function
$\Gamma(x)$ by the number of arguments.  The mass profile is
\begin{equation}
M_{\rm E}(r) = M_0 \left(\frac{\kappa}{2}\right)^{3/\kappa}
\frac{e^{2/\kappa}}{4\kappa} \, \gamma
\left[\frac{3}{\kappa},\frac{2}{\kappa} \left(\frac{r}{r_s}\right)^\kappa
  \right],
\label{Einasto mass}
\end{equation}
which yields a total mass
\begin{equation}
M_{\rm E}(\infty) = M_0 \left(\frac{\kappa }{2}\right)^{3/\kappa}
  \frac{e^{2/\kappa}}{4\kappa} \,\Gamma \left(\frac{3}{\kappa}\right),
\end{equation}
and a gravitational potential
\begin{equation}
\Phi_{\rm E}(r) = -\frac{GM_{\rm E}(\infty)}{r_s \Gamma \left(\frac{3}{\kappa}\right)} \left\{ \frac{\gamma
  \left[\frac{3}{\kappa},
    \frac{2}{\kappa}(\frac{r}{r_s})^{\kappa}\right]}{r/r_s} +
\frac{\Gamma\left[\frac{3}{\kappa},
    \frac{2}{\kappa}(\frac{r}{r_s})^{\kappa}\right]}
     {(\kappa/2)^{1/\kappa}} \right\}.
\label{Einasto potential}
\end{equation}

We adopt this profile for our spherical mass model, fixing $\kappa =
0.17$, which is the value \citet{Na04} preferred for their three
prototype galaxy groups and highest resolution haloes \citep{Na10}.

We set $r_{\max }=15$, so that the total mass of our models is
$1.497M_0$.  (Note $M_{\rm E}(r_{\max }) =1.960$ using
eq.~(\ref{Einasto mass}); the difference is due to energy truncation
that tapers the density smoothly to zero at $r_{\max }$.)  We select
particles with individual masses using $L_0=0.1$ (see \S\ref{Particle
  Resolution}) which causes $\sim 95\%$ of all particles to lie within
$r=8$, a radius we label as $r_{0.95}$.

\begin{figure*}
\begin{center}
\includegraphics[width=.28\hsize, angle=270]{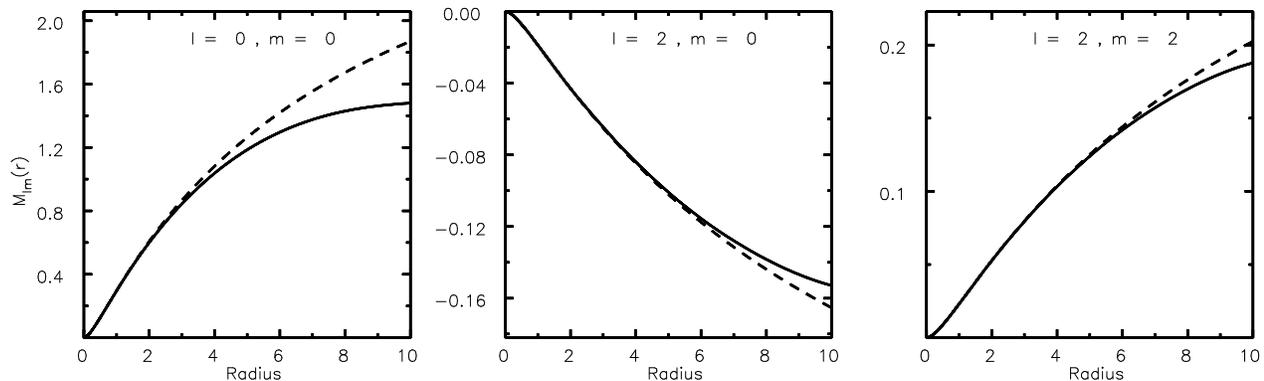}
\caption{The solid curves show the target mass harmonic profiles up to
  $l=2$ for model P.  The dashed curves show the corresponding
  profiles of an ellipsoid of density given by eq.~(\ref{Einasto
    ellipsoid}) without an energy cut off.}
\label{Mass Harmonics}
\end{center}\end{figure*}

From the discussion in \S\ref{Homoeoid}, the density of our Einasto
{\it ellipsoid} is
\begin{equation}
\rho_e(\bx) =\theta_{yz}^{-1}\rho_{\rm E}(\xi),
\label{Einasto ellipsoid}
\end{equation}
which obviously reduces to $\rho_{\rm E}(r)$ for $\varepsilon_y =
\varepsilon_z = 0$.

\begin{table}\centering
\begin{tabular}{cccc}
    &  &  &  \\ 
Label  & $(\varepsilon_{y},\varepsilon_{z}) $ & 
Axis Ratios & Triaxiality \\
\hline\hline
T$_{\rm A}$ & (0.60,0.80) & (0.80,0.60) & 0.563 \\ 
T$_{\rm B}$ & (0.70,0.80) & (0.71,0.60) & 0.766 \\ 
P          & (0.80,0.80) & (0.60,0.60) & 1.000 \\
\end{tabular}
\caption{Axis ratios and triaxiality parameters of our three Einasto
  halo models.}
\label{Model Halos}
\end{table}

\subsection{Three models}
Three ellipsoidal halo models summarized in Table \ref{Model Halos}
are the focus of this work.  All have Einasto index $\kappa =0.17$, a
semi-major axis truncated at $r_{\max} = 15r_s$, and are strongly
flattened ($c/a = 0.6$) along the minor axis, but differ in their
intermediate axes.  Although their triaxiality parameters, which are
typical of cosmologically formed halos (JS02, SFC12), increase from
model T$\rm_A$ to P, model T$\rm_A$ is the most strongly triaxial.

Ellipsoids have reflection symmetry about their three principal
planes.  We therefore align our coordinate $x$-axis with the major
axis and the $z$-axis with the polar axis in order that only the even
terms of the surface harmonic expansion are non-zero.  Oriented in
this manner, the $m=0$ components of the $l/2\geq 1$ harmonics
characterize the eccentricity along the minor axis, while even $m \geq
0$ terms describe the density along the intermediate axis.  We retain
even $l$ terms up to $l_{\max }=4$ only.

Fig.~\ref{Mass Harmonics} illustrates the profiles of the three main
contributions -- monopole and quadrupole moments -- for model P
and compares the $M_{lm}^{c}(r)$, $l\leq 2$, computed from the
smooth density $\rho_e$, with those for $\bar{M}_{lm}^{c}(r)$ for the
target ellipsoid.

\subsection{Units}
\label{sec.units}
We employ natural units, for which $r_s=M_0=G=1$, throughout.  The
unit of velocity is therefore $v_{\rm dyn} \equiv (GM/r_s)^{1/2}$ and
the dynamical time is $\tau_{\rm dyn} \equiv (r_s^3/GM)^{1/2} =
r_s/v_{\rm dyn}$.

\begin{table}\centering
\begin{tabular}{lccccc}
& Aq & SmP & T$_{\rm A}$ & T$_{\rm B}$ & P \\ \hline\hline
$r_s\;$(kpc) & $20.3$ & $20.0$ & $20.0$ & $20.0$ & $20.0$ \\ 
$r_{200}\;$(kpc) & $177.3$ & $232.0$ & $209.9$ & $210.1$ & $210.1$ \\ 
$r_{\max }\;$(kpc) & $39.5$ & $31.3$ & $33.4$ & $31.7$ & $30.7$ \\ 
$v_{\max }\;$(km s$^{-1}$) & $203.2$ & $216.3$ & $206.2$ & $210.1$ & $215.7$ \\
$c_{200}$ & $8.7$ & $11.6$ & $10.5$ & $10.5$ & $10.5$ \\ 
$\mu_{200}(\times 10^{11})$ & $12.95$ & $13.48$ & $9.97$ & $10.0$ & $10.0$
\end{tabular}
\caption{Rotation curve characteristics of halo models.  Aq refers to
  halo model Aq-D-2 in \citet{Na10}.  The column marked SmP refers to
  the smooth ellipsoid $\rho_e$ with the eccentricities of the P model}
\label{Rotation Curve}
\end{table}

\begin{figure}
\begin{center}
\includegraphics[width=.9\hsize]{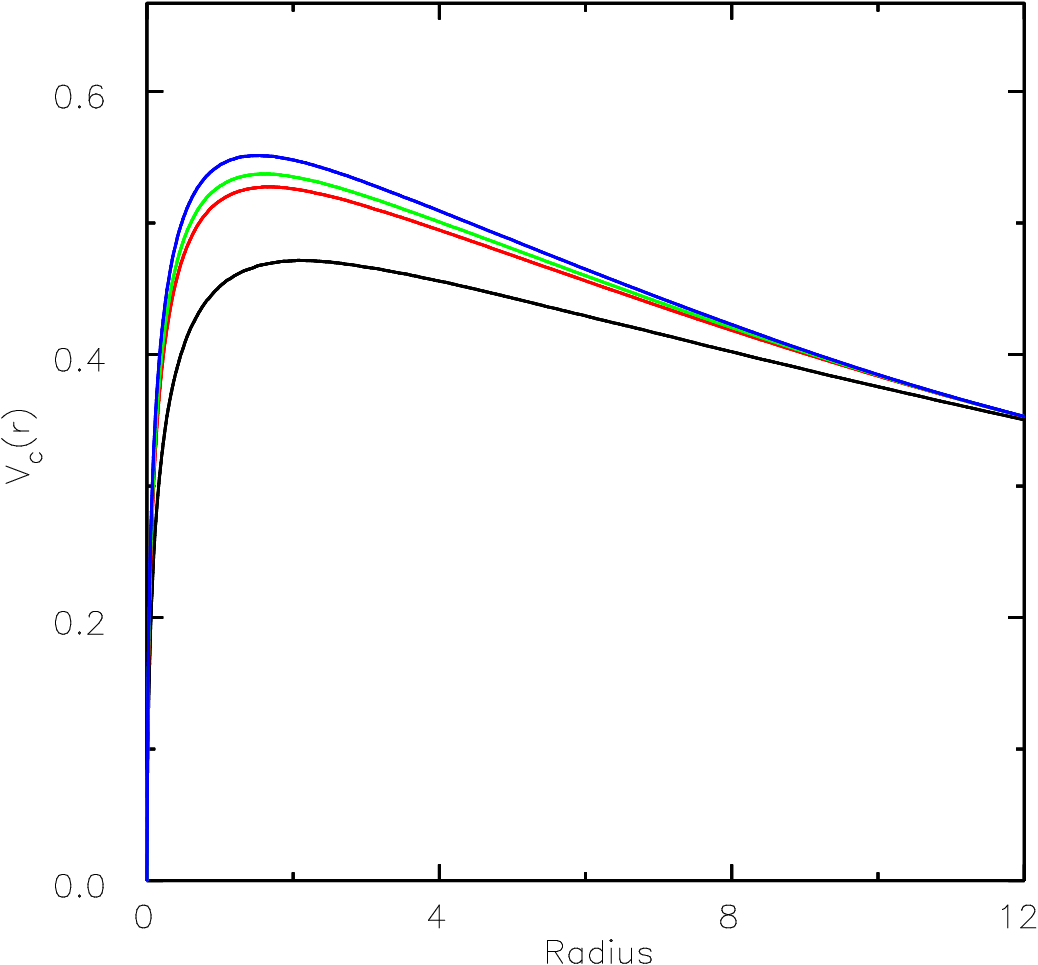}
\end{center}
\caption{Spherically averaged rotation curves $V_c(r) =
  \left[G\bar{M}_{00}^c(r)/r\right]^{1/2}$ (in natural units) for each
  model.  The lower curve (black) is that of the underlying spherical
  halo, while increasing the triaxiality parameter yields higher
  velocity peaks: T$_{\rm A}$ (red), T$_{\rm B}$ (green), and P
  (blue).}
\label{Rotation curves}
\end{figure}

We compare our models with the cosmologically-simulated halo model
Aq-D-2 extracted from \citet[][their Tables 1 and 2]{Na10}, which also
has the Einasto index ($\kappa =0.17$).  Fig.~\ref{Rotation curves}
shows rotation curves computed from the spherically averaged mass,
$v_{\rm rot}(r) = \left[\bar{M}_{00}^{c}(r)/r\right]^{1/2}$ from which
we deduce the peak circular $v_{\rm max }$ and virial $v_{200}$
velocities.  Setting the length scale to $r_s=20\,$kpc, close to that
of Aq-D-2, and $\tau_{\rm dyn} =50\,$Myr, we have $M = 6.74 \times
10^{11}\;$M$_\odot$ and velocity unit $v_{\rm dyn} \simeq 391\;$km
s$^{-1}$.  The figure highlights the impact of spherically averaging
the target density on otherwise identical models of different
eccentricities.

\subsection{Implementation}
\label{sec:implement}
Each of the halo models in Table~\ref{Model Halos} to be simulated in
an M2M run consists of $N_M = 1.2 \times 10^6$ particles, while their
associated target models -- from which we derive the target potential,
moments and dispersions -- have $N_T= 4.5 \times 10^7$ particles.

With the parameter $L_0 = 0.1$, we find $\sim 2.5\times 10^5$
particles inside a radius $r=0.1$ (first 16 grid nodes) for the
simulation models.  To obtain a specific M2M binning scheme, we sample
$N_M$ particles from the the target population then identify
aggregations of gravity grid nodes with a minimum of $2.5\times 10^5$
particles or a maximum of $10$ such nodes, except near the center
where strict adherence to the criterion makes the first few bins too
wide: we shorten the first bin to 5 nodes, which contains between 6000
to 8000 particles; conversely, at the boundary we widen the last
occupied bin to 20 to 30 nodes so that it contains $\ga 1000$
particles.  This particular scheme typically yields a maximum of $60$
M2M radial bins ($56$ of which are occupied initially) to encompass
the $501$ gravity-grid nodes.  In cases where a given target value
$\bar{M}_{k,lm}^\alpha$ vanishes due to symmetry, sampling typically
yields a nonzero mean with a dispersion that greatly exceeds the mean;
we omit the corresponding $\Delta_{k;lm}^\alpha$ when solving the FOCE
to further improve efficiency.

Step (ii) in \S \ref{Initialization} achieves the ellipsoidal
arrangement, and (iii) a phase-space near equilibrium.  The velocity
adjustments in the last step yield a mild radial anisotropy, $\beta$
(BT08, see also \S\ref{Velocities}) that gradually increases from $\ga
0$ (isotropy) at the origin to a maximum near the boundary as the
triaxiality parameter increases, i.e., $\beta \sim 0.1$ for model
T$_{\rm A}$, while $\beta \sim 0.3$ for model P.

We compute accelerations from the rigid target potential during the
M2M evolution, and advance the motion of all particles using leapfrog
with a time step $\Delta t=0.0025$.  After each time step, we solve
the FOCE using a fixed minimum number $n_F$ of iterations, gradually
increasing $n_F$ (from $n_F=5$ initially to a maximum of $12$) so that
at the end of the M2M evolution $\vert \Delta_{k;lm}^\alpha\vert <
0.1$ on average.  To achieve this threshold we choose $\mu =0.5$ and
$\varepsilon_0=0.005$.  Although the entropy in the P and T$_{\rm B}$
models converges sooner, we still iterate all models for 200 dynamical
times, in order to allow particles in the outer halo envelope to
complete more than one full orbit and to enable most off-grid
particles to return.  Only a negligible number remains permanently
outside the grid, as reported below (Table \ref{M2M weights}).

After the end of the M2M evolution, we continue the evolution for a
further 100 dynamical times with a fully self-consistent simulation.
This continued evolution enables us to check the long term survival of
the shape imposed by the M2M technique.  We also increase $l_{\max }$
to $6$, to better capture the strong flattening along the polar axis,
and include odd-$l \ (<l_{\max})$ contributions to the force
determination, as further tests of the robustness of the shape.

\begin{figure*}
{\begin{center}
\includegraphics[width=.28\hsize,angle=270]{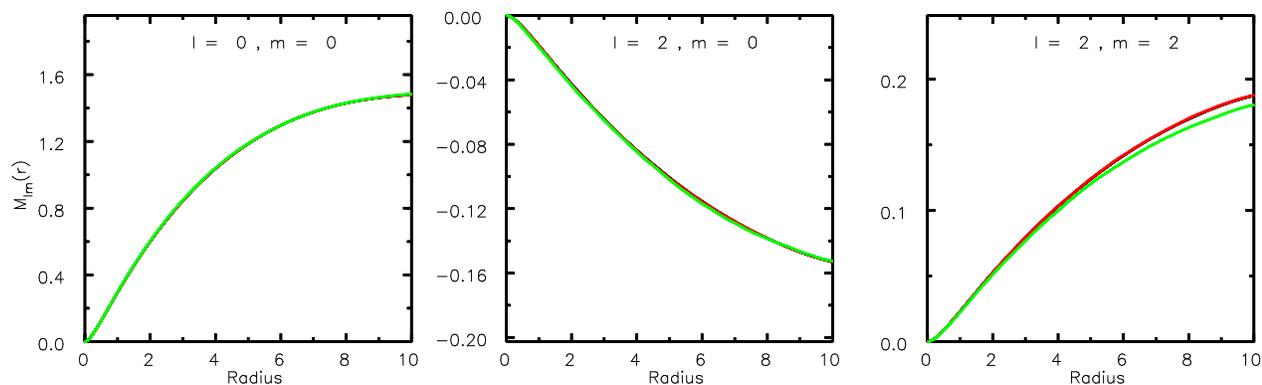}
\end{center}}
\caption{Target (black), M2M (red) and self-consistent (green) mass
  harmonic cosine curves of the model P.  The black curves are barely
  visible due to the excellent agreement of the red curves.}
\label{Mass Curves}
\end{figure*}

\section{Results}
Tables \ref{M2M statistics} and \ref{M2M weights} summarize the range
of values of key M2M quantities for our three models near or at
completion of the procedure.  Since the behavior of the
M2M objective, $\cal L$, is essentially stochastic, Table~\ref{M2M
  statistics} gives the best and worst values of the objective to
which we add the associated entropy $\cal S$, cost function
$\cal C$ and bin threshold, $\max\vert \Delta_{k;lm}^\alpha\vert$.

The more strongly elongated models have greater order, a trend that is
reflected in $\Delta S(\infty)$.  To support this interpretation, we
reran a version of model T$_{\rm A}$ -- model T$_{\rm A2}$ -- for
which we omitted steps (ii) and (iii) of the initialization, i.e.,
used a spherically symmetric halo, but otherwise used the same
parameters and procedure during the M2M evolution as for the T$_{\rm
  A}$ model.  The resulting entropy converges to the lower value
$-0.80$; hence, nearly $40\%$ of the prospective change in the entropy
is captured through the direct distortion of the spherical model in
our initialization process.

\begin{table}\centering
\begin{tabular}{ccccccc}
& \multicolumn{2}{c}{T$_{\rm A}$} & \multicolumn{2}{c}{T$_{\rm B}$} & 
\multicolumn{2}{c}{P} \\ \cline{2-7}
& B & W & B & W & B & W \\ \hline\hline
Objective (${\cal L}$) & -0.39 & -0.52 & -0.43 & -0.75 & -0.45 & 
-0.64 \\ 
Chi-square (${\cal C}$) & 0.27 & 0.53 & 0.32 & 0.96 & 0.31 & 0.68 \\ 
Entropy ($\Delta{\cal S}$) & -0.52 & -0.52 & -0.54 & -0.54 & -0.59 & -0.59 \\ 
$\max \vert \Delta_{k;lm}^{\alpha }\vert$ & 0.08 & 0.14 & 0.09 & 0.16 & 0.07 & 0.13
\end{tabular}
\caption{Best (B) and worst (W) M2M objective near completion of algorithm.
Other quantities associated with the same instances. }
\label{M2M statistics}
\end{table}

Table \ref{M2M weights} gives the percentage of off-grid particles and
of particles that have been driven to a zero weight, $N_0(\infty)$.
These are but a tiny fraction of $N_M$.  The percentage of massless
particles that lie within the radii $r_{0.95}$, $N_0(r_{0.95})$, and
$r_{0.95}/2$, $N_0(r_{0.95}/2)$, at the end of the M2M evolution
indicates that most massless particles are near the outer edge.  The
increase in the numbers of massless particles with the triaxiality
parameter is strongly correlated with the lower (greater) entropy
change (ordering), whereas the off-grid particle count, which also
appears to be correlated with triaxiality, is actually driven by the
initialization method.  Thus, there are {\it no\/} off-grid particles
for model T$_{\rm A2}$, supporting the earlier observation of the
effect of initialization steps (ii) and (iii) as tending to generate
highly eccentric outer orbits.  In this case also $N_0(\infty) \sim
0.18\%$ -- i.e., there are five times as many zero mass particles as
in T$_{\rm A}$.

\begin{table}\centering
\begin{tabular}{c|cccccc}
&  & T$_{\rm A}$ &  & T$_{\rm B}$ &  & P \\ \hline\hline
Off-grid &  & 0.002 &  & 0.005 &  & 0.025 \\ 
$N_0(\infty)$ &  & 0.033 &  & 0.075 &  & 0.097 \\ 
$N_0(r_{0.95})$ &  & 0.012 &  & 0.025 &  & 0.035 \\ 
$N_0(r_{0.95}/2)$ &  & 0.005 &  & 0.009 &  & 0.014
\end{tabular}
\caption{Percent of off-grid and zero-weighted particles at the end of
  the M2M evolution.}
\label{M2M weights}
\end{table}

Fig.~\ref{Mass Curves} shows the radial variation of
$\bar{M}_{lm}(r)$, or target profiles, and $m_{lm}(r)$, derived from
the particles, for model P.  The curves are drawn to $r=10$, which
contains $>98\%$ of the particles and at two times: the end of both
the M2M and self-consistent runs; their close agreement in the inner
halo gives an indication of how well the desired shape is achieved.

For a more quantitative measure, we give values of the {\it cumulative
  relative error}, $\delta _{lm}\equiv \vert m_{lm}(r_{.95})
-\bar{M}_{lm}(r_{.95})\vert / \vert \bar{M}_{lm}(r_{.95})\vert$ in
Table \ref{Target Deviations} for the principal $(l,m)$ contributions
explicitly included in the M2M algorithm.  Not surprisingly, the
relative errors are smallest for the monopole terms, which are
numerically the largest.  Since all three halos are equally strongly
flattened, the smallest errors are for the $(l,0)$ components, while
the errors for the $(l,m>0)$ components decrease with increasing
triaxiality parameter, a relationship that is reinforced in the
prolate model P.  This trend can also be understood in terms of the
sampling uncertainty due to the relative contributions from the higher
harmonics: decreasing the triaxiality parameter at fixed flattening
decreases the contribution of the $m>0$ harmonics and increases the
associated sampling uncertainty relative to the $m=0$ harmonics.  Note
that for the self-consistent runs, the smaller error in the $m > 0$
components as triaxiality parameter increases follows from the same
reasoning.

\begin{table}\centering
\begin{tabular}{ccccccc}
& \multicolumn{2}{c}{T$_{\rm A}$} & \multicolumn{2}{c}{T$_{\rm B}$} &
\multicolumn{2}{c}{P} \\ \hline
$(l,m)$ & M2M & SC & M2M & SC & M2M & SC \\ \hline\hline
$(0,0)$ & -3.94 & -2.76 & -3.53 & -2.87 & -3.50 & -3.15 \\ 
$(2,0)$ & -4.43 & -1.95 & -3.11 & -2.39 & -3.08 & -2.72 \\ 
$(2,2)$ & -2.27 & -1.25 & -2.40 & -1.39 & -2.48 & -1.42 \\ 
$(4,0)$ & -1.66 & -1.31 & -1.56 & -1.35 & -1.47 & -1.74 \\ 
$(4,2)$ & -2.46 & -0.61 & -2.57 & -0.89 & -3.01 & -1.02 \\ 
$(4,4)$ & -1.56 & -0.19 & -1.56 & -0.94 & -1.89 & -1.45
\end{tabular}
\caption{$Log_{10}$ of cumulative error $\delta_{lm}$ of the mass
  harmonics at end of both M2M and self-consistent (SC) runs.}
\label{Target Deviations}
\end{table}

\begin{figure*}\centering
{\begin{center}
\includegraphics[width=.9\hsize,angle=270]{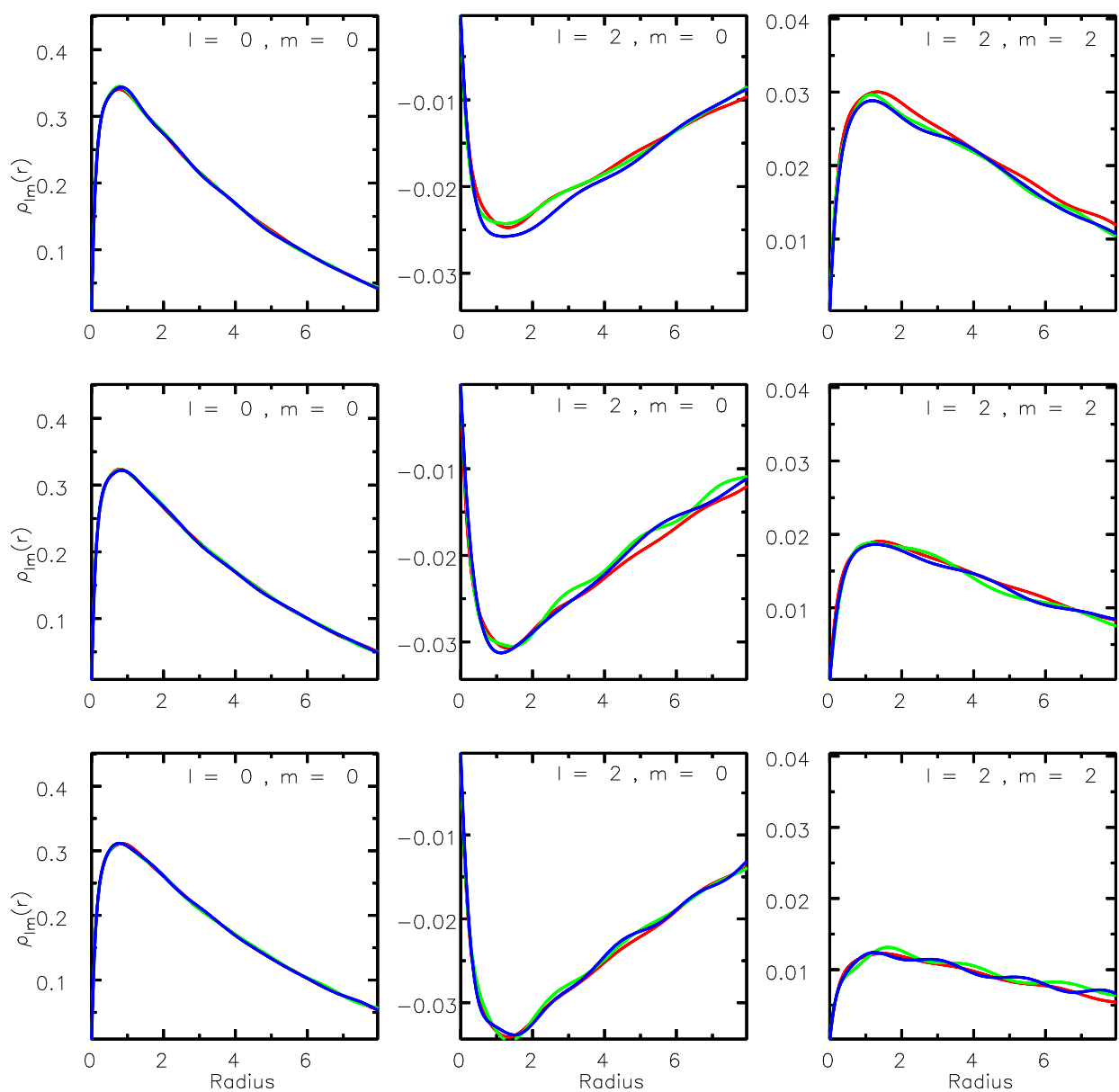}
\end{center}}
\caption{Monopole and quadrupole density moments of P (top row),
  T$_{\rm B}$ (middle row), and T$_{\rm A}$ (bottom row) models.
  Curves displayed are at the ends of the M2M (red), self-consistent
  (green) and perturbed (blue) phases.}
\label{Density Curves}
\end{figure*}

\subsection{Testing Self-Similarity}
%\subsection{Fitting the Density}
\citet{Ka91}, JS02 and \citet{De08} analyze the shapes of their halos
directly from the particles by fitting an ellipsoid over iteratively
refined particle volumes, while SFC12 diagonalizes the local inertia
tensor.

We avoid these numerically costly techniques and instead use a
least-squares fit with Chebyshev polynomials of the values of
$m_{lm}^\alpha(r_k)$ at each node $k$ of the gravity grid, but require
the value of the fit and its derivative to vanish at the origin.
Differentiation of the fitted function, $\mu_{lm}^\alpha(r)$, yields
the corresponding weighted-densities, $r^2\rho_{lm}^\alpha(r)$, shown
in Fig.~\ref{Density Curves}.  These
components are combined as in eq.~(\ref{Harmonic expansion}) to yield
an approximate expansion for the total density
\begin{eqnarray}
\rho (\br) & \approx & \sum_{l=0}^{l_{\max}} \sum_{m=0}^l \gamma_{lm}
\Pi_l^m(\theta) \big[ \cos m\phi \, \rho_{lm}^c(r) +  \nonumber \\ 
&& \qquad\qquad\qquad\qquad\qquad \sin m\phi \, \rho_{lm}^s(r) \big],
\label{Density Expansion}
\end{eqnarray}
where $\rho_{lm}^\alpha(r) = r^{-2}d\mu_{lm}^\alpha/dr$.

\begin{figure*}\centering
{\begin{center}
\includegraphics[width=.9\hsize,angle=270]{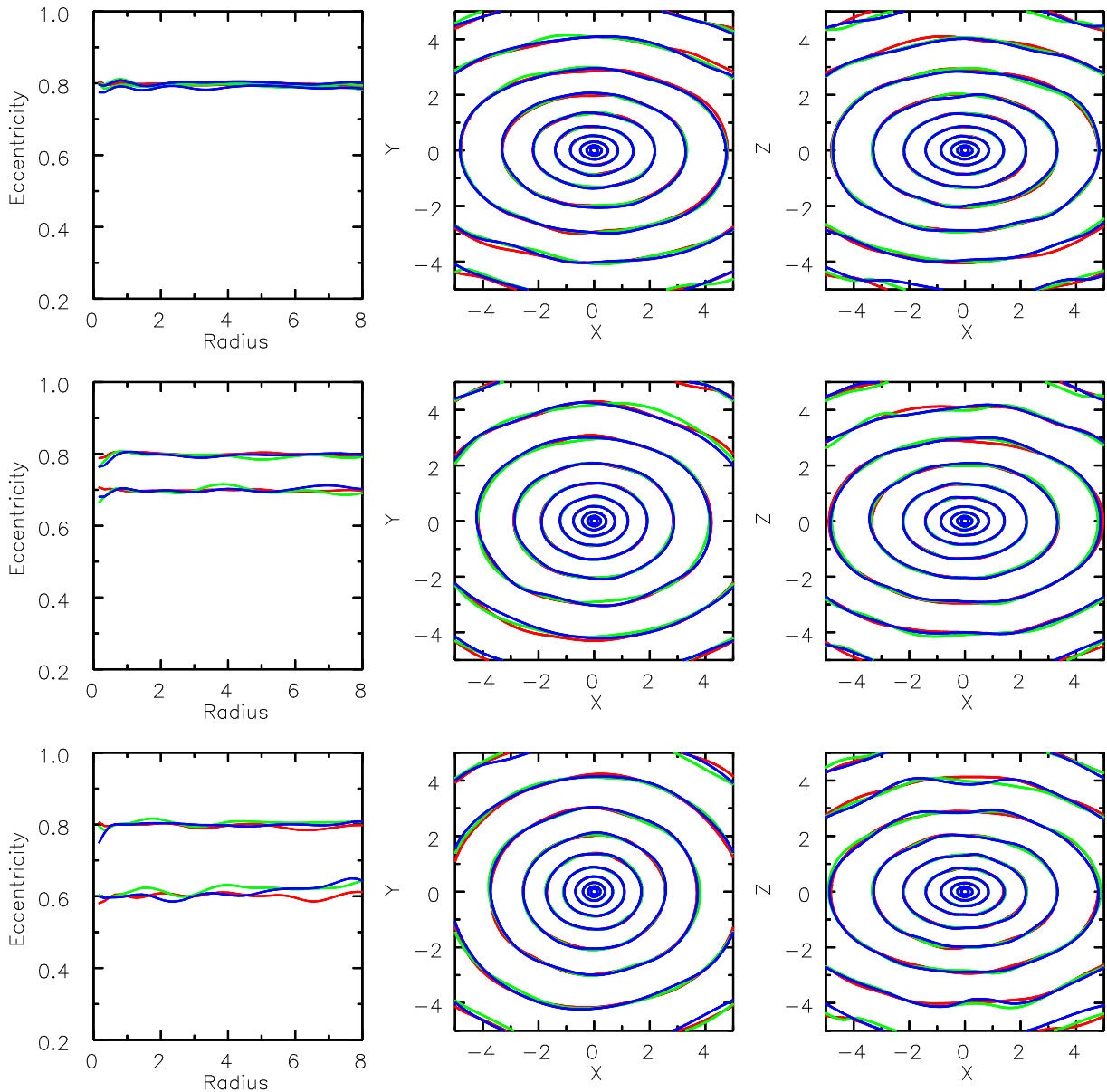}
\end{center}}
\caption{The left panels show the eccentricity profiles derived from
  least-squares ellipse fits to P (top), T$_{\rm B}$ (middle), and
  T$_{\rm A}$ (bottom) models.  The middle and right panels of each
  row show density contours on the $XY$- and $ XZ$-planes,
  respectively.  The colours in all panels refer to the ends of
  of the M2M (red), self-consistent (green) and perturbed (blue)
  evolution.}
\label{Triaxialities}
\end{figure*}

\begin{table*}\centering
\begin{tabular}{cccccccccccccccc}
& \multicolumn{1}{|c}{} & \multicolumn{4}{|c}{T$_{\rm A}$} &  & 
\multicolumn{4}{|c}{T$_{\rm B}$} &  & \multicolumn{4}{|c}{P}
\\ 
& \multicolumn{1}{|c}{} & \multicolumn{2}{c}{$\varepsilon_{y}=0.60$} & 
\multicolumn{2}{c}{$\varepsilon_{z}=0.80$} &  & \multicolumn{2}{c}{$\varepsilon
_{y}=0.70$} & \multicolumn{2}{c}{$\varepsilon_{z}=0.80$} &  & 
\multicolumn{2}{c}{$\varepsilon_{y}=0.80$} & \multicolumn{2}{c}{$\varepsilon
_{z}=0.80$} \\ \hline
$x$ &  & M2M & SC & M2M & SC &  & M2M & SC
& M2M & SC &  & M2M & SC & M2M & SC \\ 
\hline\hline
0.10 & \multicolumn{1}{|c}{} & 0.585 & 0.613 & 0.816 & 0.809 & 
& 0.703 & 0.700 & 0.798 & 0.791 &  & 0.811 & 0.808 & 
0.816 & 0.806 \\ 
0.15 & \multicolumn{1}{|c}{} & 0.581 & 0.604 & 0.808 & 0.798 & 
& 0.698 & 0.694 & 0.791 & 0.784 &  & 0.803 & 0.798 & 
0.807 & 0.800 \\ 
0.25 & \multicolumn{1}{|c}{} & 0.583 & 0.596 & 0.799 & 0.786 & 
& 0.695 & 0.689 & 0.791 & 0.779 &  & 0.796 & 0.788 & 
0.799 & 0.795 \\ 
0.50 & \multicolumn{1}{|c}{} & 0.599 & 0.598 & 0.796 & 0.793 & 
& 0.702 & 0.698 & 0.803 & 0.790 &  & 0.796 & 0.789 & 
0.798 & 0.804 \\ 
0.75 & \multicolumn{1}{|c}{} & 0.603 & 0.602 & 0.800 & 0.801 & 
& 0.703 & 0.704 & 0.805 & 0.800 &  & 0.800 & 0.796 & 
0.801 & 0.812 \\ 
1.00 & \multicolumn{1}{|c}{} & 0.599 & 0.607 & 0.801 & 0.803 & 
& 0.700 & 0.702 & 0.802 & 0.802 &  & 0.799 & 0.795 & 
0.800 & 0.808 \\ 
2.00 & \multicolumn{1}{|c}{} & 0.606 & 0.613 & 0.803 & 0.814 & 
& 0.696 & 0.698 & 0.800 & 0.798 &  & 0.799 & 0.797 & 
0.799 & 0.792 \\ 
3.00 & \multicolumn{1}{|c}{} & 0.599 & 0.602 & 0.790 & 0.794 & 
& 0.698 & 0.698 & 0.799 & 0.794 &  & 0.798 & 0.796 & 
0.794 & 0.796 \\ 
4.00 & \multicolumn{1}{|c}{} & 0.608 & 0.629 & 0.797 & 0.807 & 
& 0.701 & 0.714 & 0.798 & 0.800 &  & 0.799 & 0.798 & 
0.796 & 0.801 \\ 
5.00 & \multicolumn{1}{|c}{} & 0.599 & 0.621 & 0.800 & 0.807 & 
& 0.701 & 0.700 & 0.801 & 0.786 &  & 0.798 & 0.793 & 
0.798 & 0.797 \\ 
6.00 &  & 0.597 & 0.618 & 0.786 & 0.804 &  & 0.693 & 
0.691 & 0.799 & 0.787 &  & 0.799 & 0.790 & 0.799 & 0.792
\end{tabular}
\caption{Eccentricity profiles at ends of the M2M and self-consistent
  (SC) evolution.}
\label{Eccentricity Table}
\end{table*}

Eq.~(\ref{Density Expansion}) forms the basis for testing
self-similarity.  We examine both the radial eccentricity profiles and
planar equidensity contours of our halo models by fitting ellipses to
the contours with the least-squares algorithm of \citet{BM77}, a
procedure that allows eccentricity, orientation and centering to be
estimated simultaneously.  The left-hand column of
Fig.~\ref{Triaxialities} shows the eccentricity profile of the primary
models, while the middle and rightmost columns show, respectively, the
density contours on the $XY$- and $ZX$-planes.
Table~\ref{Eccentricity Table} gives the eccentricity values at a
fixed set of points along the major axis evaluated at the ends of both
the M2M and unconstrained evolution.  Of particular interest are those
of the self-consistent runs as they correspond to freely
evolving, equilibirum halos.  Except for mild deviations in the outer
halo, the tabulated values show remarkable uniformity along a given
principal axis.  It is noteworthy that the magnitude of these
deviations decreases with increasing eccentricity.  Thus, while the
eccentricity along the minor axis is very stable for all three models
($< 3\%$ error), along the largest departures ($\sim5\%$) on the
intermediate axis arise in the model with the least eccentricity.  The
centers of all computed ellipses differ by no more than a part in
$10^3$, while the direction of the major axis fluctuates by less than
$2^\circ$, declining outwards, with no systematic difference caused by
the unconstrained evolution.

In order to compute projected surface densities, we note that
eq.~(\ref{Self-similarity condition}) can be rewritten as
$\xi^2=\tilde q^2 + z^2/(1-\varepsilon_{z}^2)$.  Thus a projection
along the $z$-axis yields a surface density $\sigma (\tilde{q})$.  For
a projection onto the $XY$-plane, we integrate along the polar axis to
obtain a surface density expansion
\begin{equation}
\Sigma (R,\varphi) = \sum_{m=0}^{l_{\max }}\left[ \Sigma_m^c(R) \cos
  m\varphi + \Sigma_m^s(R) \sin m\varphi \right]
\label{Surface density expansion}
\end{equation}
such that
\begin{equation}
\Sigma_m^\alpha(R) = \sum_{l=m}^{l_{\max }}\gamma _{lm}
\int_{-\infty}^\infty dz \,\Pi_l^m \left(\frac{z}{r}\right)
\rho_{lm}^\alpha(r),
\label{Harmonic surface density}
\end{equation}
where $r = (R^2 + z^2)^{1/2}$.  Table~\ref{Projected Eccentricity
  Table} gives computed eccentricities for all three models for the
self-consistent runs.  To perform a similar projection along a
different axis, particle coordinates are first rotated so that the
axis of projection becomes the new polar axis, then apply the above
equations.

\subsection{Stability}
\label{sec.stability}
If there were any {\it linearly} unstable modes present in our models,
they would be seeded by shot noise.  But all three models retain close
to a constant eccentricity and axial alignment in their interior over
100 dynamical times of self-consistent evolution, hence any such modes
must have negligible growth rates.

To test for {\it nonlinear} stability, we conduct an additional
simulation in which we introduce into each of our halo models a
perturber that is turned on adiabatically over a 50 dynamical times
period and turned off over the same time period; we refer to these as
perturbed runs.  We try these experiments on the models at the end of
the respective self-consistent run.  Our adpoted perturbation is the
$l=2$ density components of an Einasto ellipsoid with the same scale,
same dimension and same $\alpha$-parameter as the host halo, but only
10\% of the mass and different eccentricities.  We use a perturber
with ($\varepsilon_y,\varepsilon_z) = (0.55, 0.85)$ for the T$_{\rm
  A}$ model, (0.65,085) for T$_{\rm B}$, and (0.75,0.85) for P
\citep[see][for a similar concept]{MB86}, and our experiments
terminate after 100 dynamical times.  In figs. ~\ref{Density Curves}
and ~\ref{Triaxialities}, the corresponding results for perturbed runs
are shown in blue.  Some quantitative differences are evident: model P
is no longer perfectly prolate and deviations from the target
self-similar profile are $<10\%$ very near the center.  Nevertheless,
the figures remain qualitatively unchanged and may be therefore deemed
as highly stable.

\subsection{Anisotropy}
\label{Velocities}
We compute the spherically-averaged anisotropy parameter $\beta (r) =
1-[\sigma_\theta^2(r) + \sigma_\varphi^2(r)] / 2\sigma_r^2(r)$ (BT08)
over the range $0 < r \leq r_{0.95}$.  Here, $\sigma_\theta^2(r)$,
$\sigma_\varphi^2(r)$ and $\sigma_r^2(r)$ are respectively the
variances of the polar, azimuthal, and radial velocities.

\begin{figure}
\begin{center}
\includegraphics[width=.9\hsize]{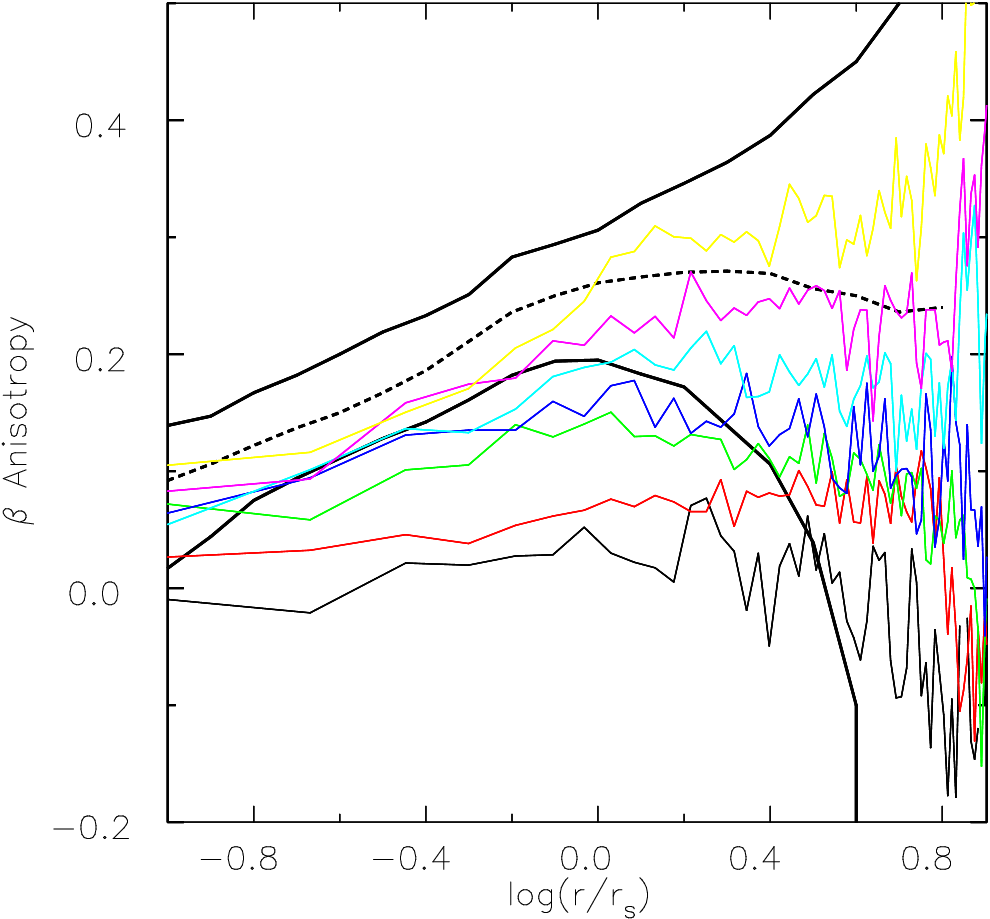}

\end{center}
\caption{ Anisotropy profiles arranged in ascending triaxiality
  parameter: Spherical (thin black line), Oblate (red), T$_{\rm D}$
  (green), T$_{\rm C}$ (blue), T$_{\rm B}$ (cyan), T$_{\rm A}$
  (magenta), and P (yellow) halo models (see text for definition of
  non-fiducial models).  Superimposed on these are the upper and lower
  boundaries (thick black lines) of the profiles in Fig.~3 of
  \citet{Lu11} and the corresponding median curve(broken black line).}
\label{Anisotropy}
\end{figure}

Fig.~\ref{Anisotropy} shows anisotropy profiles for our three
principal models at the end of their respective self-consistent runs,
but also includes four other models with lower triaxiality parameters
for comparison.  These are a sphere (S), an oblate spheroid (O) with
$\varepsilon_z = 0.8$, and two low triaxiality ellipsoids with
$(\varepsilon_y, \varepsilon_z) = (0.3,0.8)$ (T$\rm_C$) and
$(0.45,0.8)$ (T$\rm_D$).  These additional models were created in the
same manner as our principal models through the self-consistent runs,
but for the sphere we fit only the monopole component during the M2M
run and admit $l_{max} = 2$ for the self-consistent evolution.
Fig.~\ref{Anisotropy} confirms that the sphere has $\beta \approx 0$
as expected, especially in the well-resolved interior region.  The
tangential bias ($\beta <0$) near the boundary is an artifact of
enforcing the targeted monopole moments in that region; the same
effect can be seen in the oblate and triaxial models T$\rm_C$ and
T$\rm_D$.  Flattening causes a radial bias ($\beta > 0$) in the inner
regions, which is enhanced in the mildly triaxial models, gradually
overcoming the tangential bias at the boundary.  Indeed, the fiducial
models, which have the highest triaxialities, are radially biased
throughout.

The thick black lines in the same figure are adapted from the middle
panel of Fig.~3 of \citet{Lu11}, which consists of halos in the
Millennium-II simulation with Einasto parameter $\alpha = 0.178$.  The
broken black line shows the median anisotropy profile of their halo
sample, while the thick black lines represent the upper and lower
bounds of an envelope that encloses the bulk of the plotted profiles.
Profiles for our fiducial models all lie within the envelope, while
those of the added models with low triaxiality either straddle or lie
beneath the lower boundary, a result consistent with the high
triaxiality, prolate bias exhibited by cosmological halos (SFC12).
Thus, even as an idealization, the self-similar triaxial geometry is
supported by an anisotropy profile indistinguishable from that of
cosmological halos with comparable triaxiality, a noteworthy result
given that we have not constrained the velocities in our models.

\subsection{Run time}
\label{Runtime}
\begin{figure*}
\begin{center}
\includegraphics[width=.3\hsize,angle=270]{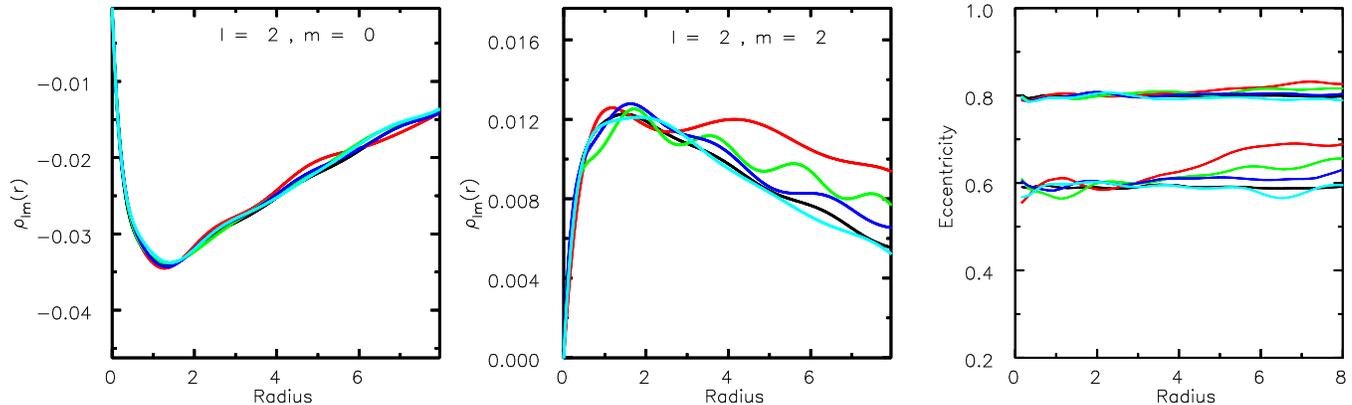}
\end{center}
\caption{Quadrupole weighted densities $m = 0$ (left), $m = 2$
  (center) and eccentricity profile of the T$\rm_A$ model for target
  (black) and M2M run times of 100 (red), 150 (green), 200 (blue) and
  250 dynamical times (cyan) shown after respective self-consistent
  runs.}
\label{Convergence Times}
\end{figure*}

In keeping with the averaging property of the FOCE (see
\S\ref{sec:FOCE}), every particle ought to complete several orbits to
ensure convergence.  Periods of orbits with radius near and beyond
$r_{200}$ of Einasto halos easily exceed 100 dynamical times so strict
adherence to such a criterion would require excessively long running
times.  However, a more reasonable length of M2M evolution achieves
robust self-similarity in the halo's interior while self-similarity in
the outer region of the halo remains less than perfect.  This is
illustrated in Fig.~\ref{Convergence Times} which displays the
quadrupole densities $\rho_{20}(r)$ (right) and $\rho_{22}(r)$
(center), and the eccentricity profile (left) after running times of
100 (red), 150 (green), 200 (blue) and 250 (cyan) dynamical times and
the subsequent self-consistent run for the T$\rm_A$ model.  We have
omitted the monopole density $\rho_{00}$ since all three cases align
perfectly with the targeted density, which suggests it is the first
moment to fully converge.  It is evident that the $\rho_{20}$ moment
converges sooner than $\rho_{22}$, especially near the boundary, which
is reflected in the eccentricity profile.  We have verified that model
P has converged adequately by 150 dynamical times.  Table \ref{Target
  Deviations} underscores the connection between accuracy of fitting
the various moments, high eccentricity and its reduction in the
uncertainty of the moment due to particle noise.  In our
implementation, more eccentric models converge more rapidly.

\begin{table}\centering
\begin{tabular}{ccccc}
& \multicolumn{1}{|c}{} & T$_{\rm A}$ & T$_{\rm B}$ & P \\ 
\hline
$x$ &  & $\varepsilon_{y}=0.60$ & $\varepsilon_{y}=0.70$ & $\varepsilon_{y}=0.80$
\\ \hline\hline
0.10 & \multicolumn{1}{|c}{} & 0.618 & 0.678 & 0.807 \\ 
0.15 & \multicolumn{1}{|c}{} & 0.601 & 0.674 & 0.795 \\ 
0.25 & \multicolumn{1}{|c}{} & 0.582 & 0.679 & 0.783 \\ 
0.50 & \multicolumn{1}{|c}{} & 0.567 & 0.701 & 0.788 \\ 
0.75 & \multicolumn{1}{|c}{} & 0.572 & 0.707 & 0.794 \\ 
1.00 & \multicolumn{1}{|c}{} & 0.591 & 0.703 & 0.793 \\ 
2.00 & \multicolumn{1}{|c}{} & 0.611 & 0.699 & 0.787 \\ 
3.00 & \multicolumn{1}{|c}{} & 0.600 & 0.696 & 0.786 \\ 
4.00 & \multicolumn{1}{|c}{} & 0.612 & 0.688 & 0.796 \\ 
5.00 & \multicolumn{1}{|c}{} & 0.611 & 0.680 & 0.790 \\ 
6.00 &  & 0.614 & 0.687 & 0.790
\end{tabular}
\caption{Profiles of the eccentricity of the surface density projected
  onto the XY-plane at end of the self-consistent runs.}
\label{Projected Eccentricity Table}
\end{table}

\section{Summary and Conclusions}
We have used the M2M method to create collisionless halo models with
self-similar, ellipsoidal geometry and a mass density profile relevant
to cosmological halos.  Our model halos are self-similar, which is not
in strict accord with cosmological results (JS02, SFC12), but is a
reasonable approximation with theoretically advantageous properties.
They are the first step in an ongoing study of the collisionless
growth of stellar disks within triaxial halos.

Both DL07 and D09 have previously created triaxial halos by this
method, but there are minor differences in our approach, and we have
adopted a more cosmologically motivated model.  The main
difference in methodology is that we do not implement a moving average,
(eq.~\ref{ODE for averaged deltas}), which we find actually degrades the
results and we show why it is superfluous.  Other minor differences in
our approach are described in \S\ref{sec:implement}.

DL07 introduced the M2M objective as a goodness-of-fit measure to
accommodate observational data.  Our present application is to a
theoretical target model, and we highlight the similarities and
differences between M2M and Schwarzschild's approach.  We also follow
DL07 and D09 by interpreting the dispersions as sampling error
consistent with shot noise inherent in any $N$-body representation,
but construct our target potential from a static $N$-body model with a
very large number of particles, which we use not only to compute the
target potential, but also to determine the mean and dispersions that
serve as proxies for $P_j$ and $\sigma_j $, respectively, in
eq.~(\ref{chi-squared statistic}).

The full procedure consists of several distinct steps: after
initializing the particles as described in \S\ref{Initialization}, we
allow the model to relax in the smooth target potential for 25
dynamical times.  We then apply the FOCE over 200 dynamical times, a
sufficient time span to ensure that changes in the entropy level are
minimal.

Our adpoted model is the Einasto halo and we present three aspherical
models that bear strong similarities to halos from the Millenium-II
simulation selected by \citet{Lu11}.  We demostrate that they are
stable by evolving them in their self-consistent potential over a
period of 100 dynamical times, during which their properties barely
changed.  We also tested non-linear stability by perturbing the model
and showed that its shape was restored quite well after the
perturbation was removed.  In additional experiments, not presented
here, we have found that singular mass profiles, such as the NFW or
Hernquist, present no difficulty, and higher eccentricities can also
be achieved, with the proviso that $l_{\max }$ may need to be
increased, which would lengthen computation times.

\section*{Acknowledgments}
We thank the referee for a careful read of the manuscript.  This work
was supported by NSF grant AST/1108977.

\def\Omit#1{{ \etal}}

\label{lastpage}

\end{document}